\documentclass[a4paper,12pt]{article}
\begin{document}
\baselineskip 8mm
\title{Symmetry Induced Heteroclinic Cycles in a $CO_2$ Laser}
\author{
R. L\'{o}pez-Ruiz$^{\ddag\S}$ and S. Boccaletti$\dag$ \\ 
{\small $^\ddag$ Dpto. de Inform\'atica e Ingenier\'{\i}a de Sistemas} \\
{\small $^\S$ Instituto de Biocomputaci\'on y F\'{\i}sica de Sistemas Complejos} \\
{\small Facultad de Ciencias, Universidad de Zaragoza, 50009-Zaragoza (Spain)} \\
{\small $\dag$ Istituto Nationale di Ottica Applicata} \\ {\small
Largo Enrico Fermi, 6, 50125-Firenze (Italy) }
\date{ }}

\maketitle
\begin{center} {\bf Abstract} \end{center}
The conditions for the existence of heteroclinic connections
between the transverse modes of a $CO_2$ laser whose setup has a
perfect cylindrical symmetry are discussed by symmetry arguments
for the cases of three, four and five interacting modes. Explicit
conditions for the parameters are derived, which can guide
observation of such phenomena.

\newpage
\section{Introduction}
Recent evidence of space-time complexity in lasers suggests the
possibility of using them as "test-benches" for space time chaos
theories [Arecchi {\it et al.}, 1990,1991,1999]\newline
\cite{dangelo,chile,green}. 
Most lasers used in applications operate with a single Gaussian
transverse mode. However, several laboratory configurations dealt
with many transverse modes simultaneously active within the
optical cavity. The nonlinear coupling between them gives rise to
a complicated dynamics, and there is hope that a careful study of
the transition from a situation where only few modes are active to
a situation with many active modes will shed light on the
mechanisms of transition to space time chaotic states. High
accuracy and short times involved in laser phenomena give to
lasers a relevant role in the study of such kind of processes.
Presently, there is a wide spectrum of experimental features
observed in transverse laser patterns waiting for theoretical
explanation. These range from configurations in which only a few
modes are active to ones in which the interaction among many modes
leads to a loss of space and time correlation. An example in the
first group appears in a laser system with photorefractive
materials \cite{arecchi}. There, the competition between different
transverse modes can give rise to periodic or chaotic alternation
between them. This phenomenon has been theoretically explained
\cite{abmg} by symmetry arguments pointing to the existence of
heteroclinic connections involving the first three transverse
modes, with the assumption that a subcritical bifurcation is
responsible for the birth of the fundamental Gaussian mode.

In this paper we consider explicitely a laser system in which the
transverse modes are all born supercritically. Our main aim is to
discuss the possibility for obtaining heteroclinic cycles when
three, four or five modes are active in a $CO_2$ laser. The paper
is organized as follows. In section 2 the existence of
heteroclinic connections in dynamical systems with symmetry is
reviewed. In section 3, the particular laser system under study is
described. The analysis of the mode equations for this system for
a perfect symmetry is done in section 4. The conditions that the
coefficients of those equations must obey in order to have some
kind of heteroclinic connection in a $CO_2$ laser system are
obtained. The main conclusions and a short discussion on possible
experiments are gathered in section 5.

\section{Symmetry Induced Heteroclinic Cycles}
Let $P_1$, $P_2$,..., $P_n$ be fixed points of a system of ODEs.
A heteroclinic cycle (HC) is a set of trajectories $Q_1(t)$,$Q_2(t)$, ...,
$Q_n(t)$ such that
\begin{eqnarray}
\lim_{t \rightarrow \infty}Q_k & = & P_{k+1}, \label{eq1}\\
\lim_{t \rightarrow - \infty}Q_k & = & P_{k}, \label{eq2}\\
P_{n+1} & = & P_1. \label{eq3}
\end{eqnarray}
Such a connection between invariant sets is non generic
for systems of ODEs. But if the system has some symmetries, it has been shown
that HCs can be structurally stable \cite{field,melbourne}.
The basic idea is that in the presence of symmetries, there are
subspaces invariant under the flow. The role of
these flow invariant subspaces in building HCs is determinant.
Take, for instance, a particular example relevant for
the system discussed in this paper.
Let
\begin{equation}
\dot{x}=f(x)
\end{equation}
be a three dimensional dynamical system with $x \in R^3$, $\Pi_i$,
$i=1,2,3$, three invariant planes and $Q_i(t)$, $i=1,2,3$, three
trajectories connecting the invariant planes (see Figure 1).
Suppose that the saddle fixed points $P_i$, $i=1,2,3,$ are the
asymptotic limits of the trajectories $Q_i(t)$ such as indicated
in the relations (\ref{eq1}-\ref{eq2}). Thus, the unstable set of
the fixed point $P_i$ is contained in the invariant plane $\Pi_i$
and its stable set is contained in the invariant plane
$\Pi_{i-1}$. Therefore, we have a heteroclinic connection between
the points $P_i$, $i=1,2,3$, forced, in this case, by the
symmetries of the system and the boundedness of the trajectories.
Let us assume that the planes are induced by subgroups of the
symmetry group of the problem. If the system is perturbed with a
term equivariant under the same group, it will have the same
invariant planes, and, as saddle-sink connections are structurally
stable in $R^2$, the cycle will persist. The situation is quite
different if the system is perturbed with a non equivariant term.
In that case, if the flow invariant planes are destroyed, one can
break the heteroclinic cycle and induce the appearance of periodic
orbits, giving rise to the phenomenon of periodic alternation, as
it has been recently shown, for instance, in a rotating convection
system \cite{lin}. We are interested in the possibility of a
periodic alternation among modes with different angular momenta
for a $CO_2$ laser in a Fabry-Perot cavity. In a perfect setup,
the boundary conditions of this system (cylindrical tube) imposes
an $O(2)$ symmetry. It has been shown that slight imperfections in
the system (a non perfect coaxial discharge, slight disalignments
of the mirrors, etc.) can break the rotational component of
$O(2)$, leaving just the $Z_2$ component (reflections)
\cite{dangelo}[L\'opez-Ruiz {\it et al.}, 1993,1994]. 
The existence of heteroclinic cycles
connecting invariant sets of a $O(2)$ equivariant system was
studied by Melbourne {\it et al.} \cite{melbourne}. Those cycles
were found in the truncated normal forms of Hopf-Hopf mode
interactions. In the following section the normal form describing
the interactions between modes in the $CO_2$ laser system is
presented and the relationships among the coefficients to have
heteroclinic cycles are determined.

\section{The System}
\label{sec3}
The system under study is a $CO_2$ laser in a Fabry-Perot cavity
\cite{solari}. The cavity
has a perfectly reflecting plane mirror at one end (z=0) and a curved
mirror with partial reflectivity at the other (z=-1). Physically, the effective
curvature of this mirror can be modified inserting a passive optical device.
As the pumping strength is changed, different modes will be excited.
The resulting dynamics can be studied by reducing the Maxwell-Bloch equations
to evolution equations for the amplitudes of the active modes. This
reduction procedure is made in two steps. First, the electric field,
the polarization and the population inversion are expanded
in terms of left and right longitudinal modes $\phi^+$, $\phi^-$,
\begin{eqnarray}
E & = & (E) e^{iwt}(e^{-\chi z} \phi^+ \Phi_+ +e^{\chi z} \phi^- \Phi_-)
(s^2+z^2)^{1/2}, \\
P & = & i(P) e^{iwt} \sum [(\phi^+)^n \Pi_n + (\phi^-)^n \Pi_{-n}]
(s^2+z^2)^{1/2}, \\
D & = & (D) (D_0+\sum [D_n (\phi^+)^n+ D_{-n} (\phi^-)^n]),
\end{eqnarray}
where $(E), (P), (D)$ stand for scale factors, $\chi$ is the rate
of cavity losses, {\em s} is a constant that depends on the
curvature of the mirrors and $z$ indicates the axial direction.
Evolution equations for the slowly varying amplitudes in that expansion
are derived. They read
\begin{eqnarray}
i\frac{\partial}{\partial t}\left[ \begin{array}{c}
\Phi_+ \\
\Phi_-
\end{array} \right] & = & \left[ \begin{array}{cc}
H^+ +i\chi & 0 \\
0 & H^- -i\chi
\end{array}\right] \left[ \begin{array}{c}
\Phi_+ \\
\Phi_-
\end{array}\right] -i\left[ \begin{array}{c}
\Pi_+ \\
\Pi_-
\end{array} \right], \label{mb1}\\
\left[ \begin{array}{c}
\Pi_+\\
\Pi_-
\end{array} \right] & = &
\left[ \begin{array}{cc}
D_0 & D_2 \\
D_{-2} & D_0
\end{array} \right] \left[ \begin{array}{c}
R^+ \Phi_+\\
R^- \Phi_-
\end{array} \right], \\
I_2+1/\gamma \frac{\partial}{\partial t} \left[ \begin{array}{cc}
D_0 & D_2\\
D_{-2} & D_0
\end{array} \right] & = &
K I_2 + \left\{ \left[ \begin{array}{c}
\Phi_+\\
\Phi_- \end{array} \right]
\left[ \begin{array}{cc}
\Pi_+^* & \Pi_-^* \end{array} \right] \right.+ \nonumber \\
& & +\left[ \begin{array}{c}
\Pi_+\\
\Pi_- \end{array} \right]
\left[ \begin{array}{cc}
\Phi_+^* & \Phi_-^* \end{array} \right] + \\
& & + \left. \left[ \begin{array}{cc}
\Phi_- \Pi_-^* +\Phi_-^* \Pi_- & 0 \\
0 & \Phi_+ \Pi_+^* +\Phi_+^* \Pi_+
\end{array} \right] \right\} (s^2+z^2)^{-1}, \nonumber \\
R^{+-} & = & (\beta-i H^{+-})^{-1}, \label{mb2}
\end{eqnarray}
where $\gamma$ is the rate of decay of the population inversion, $\beta$
is the atomic rate of decay of the atomic polarization, $K$ is the pumping
profile, $H^{+-}$ are 3D differential operators and $I_2$ is the 2D identity
matrix.
The trivial solution of these equations is
\begin{eqnarray}
\Phi_{+-} & = & 0, \\
\Pi_{+-} & = & 0, \\
D_{+-2} & = & 0,
\end{eqnarray}
and
\begin{equation}
D_0 = K.
\end{equation}
For certain values of the parameters, different cavity modes are born from a
Hopf bifurcation of this trivial solution. The electric
field amplitude can be expanded as a combination of those cavity
mode amplitudes:
\begin{equation}
\left[ \begin{array}{c}
\Phi_+ \\
\Phi_-
\end{array} \right] = \sum z_{\mu}
\left[ \begin{array}{c}
a^{\mu}_+ \\
a^{\mu}_-
\end{array} \right].
\end{equation}
Introducing this expansion into the Maxwell-Bloch equations
(\ref{mb1}-\ref{mb2}) one arrives to a set of evolution equations for
the cavity mode amplitudes:
\begin{equation}
\dot{z}_{\alpha}=L_{\alpha \beta}z_{\beta}+ M_{\alpha \mu \nu \beta}
z_{\mu}z_{\nu}^* z_{\beta}+ h.o.t. \label{eq:gen}
\end{equation}
The coefficients in these equations can be explicitly computed:
\begin{eqnarray}
L_{\alpha \beta} & = & \delta_{\alpha \beta}(
(-\chi+\frac{K\beta}{\beta^2+\Omega_{\alpha}^2})
-i\Omega_{\alpha}(1-\frac{K}{\beta^2+\Omega_{\alpha}^2})),\\
M_{\alpha \mu \nu \beta} & = & -G_{\alpha \mu \nu \beta}D_{\alpha \mu
\nu \beta},\\
D_{\alpha \mu \nu \beta} & = & \frac{1}{1+i(\Omega_{\mu}-\Omega_{\nu})/\gamma}
(\frac{1}{\beta-i\Omega_{\mu}}+(\frac{1}{\beta-i\Omega_{\nu}})^*)\times \\
& &\times \frac{1}{\beta-i\Omega_{\beta}}, \nonumber \\
G_{\alpha \mu \nu \beta}& =& \int
((a^{\alpha *}_+a^{\mu}_+ +a^{\alpha *}_-a^{\mu}_- )
(a^{\nu *}_+a^{\beta}_+ +a^{\nu *}_-a^{\beta}_- )+\\
& &+(a^{\alpha *}_+a^{\mu}_+a^{\nu *}_+a^{\beta}_+ +a^{\alpha *}_-a^{\mu}_-
a^{\nu *}_-a^{\beta}_-))\frac{K}{s^2+z^2}dv, \nonumber
\end{eqnarray}
with $\Omega_{\alpha}$ the slow temporal frequency of the empty
cavity modes. \newline
In the following section the relationships between the
coefficients of these equations to have a
HC are derived.

\section{Solutions of the Amplitude Equations}
Now the conditions for the existence of heteroclinic solutions in
the $CO_2$ laser system are discussed. If the Gaussian mode
interacts with the modes of angular momenta $\pm 1$, HCs are not
possible. This result is generic for $O(2)$ equivariant systems
when all modes are born supercritically from the origin. Then one
must go one step further considering four modes (of angular
momentum $\pm 1$, $\pm 2$) interacting in the cavity. It is
interesting to analyze this case because this is the simplest
situation in which heteroclinic cycles can be built generically
for $O(2)$ equivariant systems. However, due to the specific form
of the physical equations we prove that HC are not possible in a
$CO_2$ laser in this case. The next step is to consider five modes
with angular momenta $0$, $\pm 1$, $\pm 2$. The restrictions among
the coefficients of the normal form to have a HC are determined.
We separate the discussion of each case in separated subsections.
\subsection{Three Mode Interaction}
\label{sec4.1}
The equations describing the interaction between the modes
with angular momenta $0$, $\pm 1$ near the Hopf-Hopf bifurcation (the parameter
value for which the primary branches in which the modes are born cross
each other) are:
\begin{eqnarray}
\dot{z}_1 & = & \lambda_1 z_1 - (A(z_1z_1^*+2z_2z_2^*)+Dz_0z_0^*)z_1,\\
\dot{z}_2 & = & \lambda_1 z_2 - (A(2z_1z_1^*+z_2z_2^*)+Dz_0z_0^*)z_2,\\
\dot{z}_0 & = & \lambda_0 z_0 - (E(z_1z_1^*+z_2z_2^*)+ Bz_0z_0^*)z_0.
\end{eqnarray}
\noindent The coefficients $A$, $B$, $E$, $D$ have been explicitly
computed in the previous section. Setting $z_i= \rho_i e^{i
\phi_i}$, the amplitude equations read
\begin{eqnarray}
\dot{\rho}_1 & = & \lambda_1^r \rho_1 -(A^r(\rho_1^2 +
2\rho_2^2)+D^r\rho_0^2)\rho_1,\label{ro1}\\
\dot{\rho}_2 & = & \lambda_1^r \rho_2 - (A^r(2\rho_1^2+
\rho_2^2)+D^r\rho_0^2)\rho_2,\\
\dot{\rho}_0 & = & \lambda_0^r \rho_0 - (E^r(\rho_1^2+
\rho_2^2)+ B^r\rho_0^2)\rho_0. \label{ro2}
\end{eqnarray}
The codimension two bifurcation diagram,
$(\lambda_0^r,\lambda_1^r)$, for these equations can be found in
figure 1 in Ref. \cite{lopez}. In that figure, projections into
the amplitude subspaces are shown for different regions of the
parameter space. Only the real part of the amplitudes must be
considered because the dynamics of the phases is trivial. Notice
that the dimension of the amplitude space is three, that there are
three invariant planes (each defined by $\rho_i$=0) and one fixed
point in each of the three invariant lines. But heteroclinic
connections cannot be established between these fixed points due
to the residual symmetry that remains in the plane
$(\rho_1,\rho_2)$ after projecting in the amplitude space. This
symmetry is given by the invariance of the equations
(\ref{ro1}-\ref{ro2}) under the variable change: $(\rho_0, \rho_1,
\rho_2)$ $\rightarrow$ $(\rho_0, \rho_2, \rho_1)$. If the flow is
projected on the plane $(\rho_1,\rho_2)$ the dynamics derives
towards one of the modes, $\rho_1$ or $\rho_2$ (or $\rho_1=\rho_2$
if these amplitudes were the same at the beginning). In this
situation one of theses modes interacts with $\rho_0$ and there is
no possibility to construct a HC because the origin is a repulsor
and there are only two degrees of freedom (fig. 2).

If one assumes a subcritical bifurcation for birth of the
fundamental mode ($\lambda_0^r < 0$ in Eqs. (\ref{ro1}-\ref{ro2}),
this allows to obtain heteroclinic connections between the three
transverse modes considered here \cite{abmg}.  Since however we
are interested to study the $CO_2$ laser, this cannot be assumed;
on the contrary, all modes are born supercritically from the
origin and thus all $\lambda$'s have a positive real part.
\subsection{Four Mode Interaction}
Once excluded the interaction among three modes that bifurcate
supercritically we analyse the possibility of a heteroclinic connection
among four modes with angular momenta $\pm 1$, $\pm 2$. \par
According to Eq. (\ref{eq:gen}), the normal form describing the interaction
between four modes of angular momenta $\pm 1$, $\pm 2$ are
\begin{eqnarray}
\dot{z}_1 & = & \lambda_1z_1-(A(z_1z_1^*+2z_2z_2^*)+B(z_3z_3^*+z_4z_4^*))z_1,\\
\dot{z}_2 & = & \lambda_1z_2-(A(2z_1z_1^*+z_2z_2^*)+B(z_3z_3^*+z_4z_4^*))z_2,\\
\dot{z}_3 & = & \lambda_2z_3-(D(z_3z_3^*+2z_4z_4^*)+C(z_1z_1^*+z_2z_2^*))z_3,\\
\dot{z}_4 & = & \lambda_1z_4-(D(2z_3z_3^*+z_4z_4^*)+C(z_3z_3^*+z_4z_4^*))z_4.
\end{eqnarray}
\noindent The behaviour of this system can be analyzed setting
$z_i= \rho_i e^{i \phi_i}$ and writing for the real part of the
amplitudes the following equations,
\begin{eqnarray}
\dot{\rho}_1 & = & \lambda_1^r\rho_1-(A^r(\rho_1^2+2\rho_2^2)+
B^r(\rho_3^2+\rho_4^2))\rho_1,\label{eqn1}\\
\dot{\rho}_2 & = & \lambda_1^r\rho_2-(A^r(2\rho_1^2+\rho_2^2)+
B^r(\rho_3^2+\rho_4^2))\rho_2,\\
\dot{\rho}_3 & = & \lambda_2^r\rho_3-(C^r(\rho_3^2+2\rho_4^2)+
D^r(\rho_1^2+\rho_2^2))\rho_3,\\
\dot{\rho}_4 & = & \lambda_1^r\rho_4-(C^r(2\rho_3^2+\rho_4^2)+
D^r(\rho_3^2+\rho_4^2))\rho_4. \label{eqn2}
\end{eqnarray}
Let us observe the residual symmetry in the amplitude space:
$(\rho_1, \rho_2, \rho_3, \rho_4) \rightarrow (\rho_2, \rho_1,
\rho_4, \rho_3)$.
The same line of reasoning of the preceding subsection
can be followed: the symmetry in the plane
$(\rho_1, \rho_2) \rightarrow (\rho_2, \rho_1)$ reduces the dynamics
to one of these modes. But the interaction of this mode with
$(\rho_3, \rho_4)$ still presents the residual symmetry
$(\rho_i, \rho_3, \rho_4) \rightarrow (\rho_i, \rho_4, \rho_3)$.
Returning to the case of
three interacting modes discussed in section \ref{sec4.1} we also conclude
that there is no way to construct a heteroclinic cycle among these modes.
Let us remark that symmetry in these equations is crucial in avoiding
a heteroclinic connection.
Thus the dynamics allows to connect the plane $(\rho_1, \rho_2)$ with the
plane ($\rho_3, \rho_4$), but it dies there
without any possibility to return to the first level. Therefore, at least,
another mode must be active
in order to allow the system to come back to the initial level (in some
sense a 'pump' mode). In this form a heteroclinic
cycle could be obtained with a minimun of five modes.
\subsection{Five Mode Interaction}
Thus, we analyse the interaction between modes with angular
momenta $0$, $\pm 1$, $\pm 2$ in a system with perfect $O(2)$
symmetry. In this case symmetries could induce heteroclinic
cycles. The bifurcation equations for this case read as follow:
\begin{eqnarray}
\dot{z}_0 & = & \lambda_0z_0-(A(z_1z_1^*+z_2z_2^*)+B(z_3z_3^*+z_4z_4^*)
+Cz_0z_0^*)z_0,\label{eqfive1}\\
\dot{z}_1 & = & \lambda_1z_1-(D(z_1z_1^*+2z_2z_2^*)+F(z_3z_3^*+z_4z_4^*)
+Ez_0z_0^*)z_1,\\
\dot{z}_2 & = & \lambda_1z_2-(D(2z_1z_1^*+z_2z_2^*)+F(z_3z_3^*+z_4z_4^*)
+Ez_0z_0^*)z_2,\\
\dot{z}_3 & = & \lambda_2z_3-(G(z_3z_3^*+2z_4z_4^*)+I(z_1z_1^*+z_2z_2^*)
+Hz_0z_0^*)z_3,\\
\dot{z}_4 & = & \lambda_2z_3-(G(2z_3z_3^*+z_4z_4^*)+I(z_1z_1^*+z_2z_2^*)
+Hz_0z_0^*)z_4.\label{eqfive2}
\end{eqnarray}
\noindent A complete study of this set of equations is beyond the
scope of this paper. We focus our attention to the possibility of
obtaining the situation displayed in figure 3. \noindent Let us
first find the restrictions in the coefficients to have a
connection between the rotating waves (RW) of two interacting
modes (Fig. 4):
\begin{eqnarray}
\dot{z}_1 & = & \lambda_1z_1-(A_1z_1z_1^*+A_1'z_2z_2^*)z_1,\\
\dot{z}_2 & = & \lambda_2z_2-(A_2z_2z_2^*+A_2'z_1z_1^*)z_2.
\end{eqnarray}
After projecting in the amplitudes space ($z_i= \rho_i e^{i \phi_i}$)
these lead to:
\begin{eqnarray}
\dot{\rho}_1 & = & \lambda_1^r\rho_1-A_{1R}\rho_1^3-A_{1R}'\rho_2^2\rho_1,\\
\dot{\rho}_2 & = & \lambda_2^r\rho_2-A_{2R}\rho_2^3-A_{2R}'\rho_1^2\rho_2.
\end{eqnarray}
\noindent To obtain a connection as in Fig. 4 the unstable
manifold of $RW_1 \equiv (\sqrt\frac{\lambda_1^r}{A_{1R}}, 0)$
must coincide with the stable manifold of $RW_2 \equiv (0,
\sqrt\frac{\lambda_2^r}{A_{2R}})$. This is reached by imposing
that the eigenvalues of the Jacobian matrix are one positive (in
$RW_1$) and another negative (in $RW_2$). As the flow is bounded
and there exist no fixed point in the interior of the plane
$(\rho_1, \rho_2)$, the unstable manifold of $RW_1$ is the stable
manifold of $RW_2$. The following conditions must hold:
\begin{eqnarray}
\frac{\lambda_2^r}{\lambda_1^r} & > & \frac{A_{2R}'}{A_{1R}}, \label{eqcond}\\
\frac{\lambda_1^r}{\lambda_2^r} & < & \frac{A_{1R}'}{A_{2R}}.
\end{eqnarray}
These considerations are now extended to the five mode interaction in order
to find the conditions to have a global heteroclinic connection
of the type (Fig. 5):
\begin{eqnarray}
\rho_0 \rightarrow plane(\rho_3, \rho_4) \rightarrow
plane(\rho_1, \rho_2) \rightarrow \rho_0
\end{eqnarray}
These conditions are:
\begin{eqnarray}
\rho_0 \rightarrow (\rho_3,\rho_4) &
\left\{\begin{array}{rcl}
\frac{\lambda_2^r}{\lambda_0^r} & > & \frac{H^r}{C^r} \vspace{6mm} \\
\frac{\lambda_0^r}{\lambda_2^r} & < & \frac{B^r}{G^r}
\end{array} \right. & \label{condition1} \\
(\rho_3,\rho_4) \rightarrow (\rho_1,\rho_2) &
\left \{ \begin{array}{rcl}
\frac{\lambda_1^r}{\lambda_2^r} & > & \frac{F^r}{G^r} \vspace{6mm} \\
\frac{\lambda_2^r}{\lambda_1^r} & < & \frac{I^r}{D^r}
\end{array} \right. & \\
(\rho_1,\rho_2) \rightarrow \rho_0 &
\left\{ \begin{array}{rcl}
\frac{\lambda_0^r}{\lambda_1^r} & > & \frac{A^r}{D^r} \vspace{6mm} \\
\frac{\lambda_1^r}{\lambda_0^r} & < & \frac{E^r}{C^r}
\end{array} \right. & \label{condition2}
\end{eqnarray}
\par
Two different situations can be distinguished. The first one corresponds
to $A^r<0, H^r<0 $ and $F^r<0$. In such a case, condition (\ref{eqcond}) is
automatically verified for the three connections and those relationships
(\ref{condition1}-\ref{condition2}) reduce simply to
\begin{eqnarray}
1< \frac{B^r}{G^r} \frac{I^r}{D^r} \frac{E^r}{C^r}.
\end{eqnarray}
Let us call this case "strong" HC among
transverse modes. The physical meaning of this case clearly appears looking at
Eqs. (\ref{eqfive1}-\ref{eqfive2}). Indeed, selecting $A^r<0, H^r<0$ and $F^r<0$,
any growth of $\rho_0$ produces a source term in the
equations for ($\rho_3, \rho_4$); any growth of ($\rho_3, \rho_4$) produces
a source term in the ($\rho_1, \rho_2$) equations and any growth of
($\rho_1, \rho_2$) gives rise to a source term in the $\rho_0$ equation.
So the cycle can close itself. \par
Guided by such a suggestion, the values of nonlinear coefficients
of Eqs. (\ref{eqfive1}-\ref{eqfive2}) are calculated directly
from Maxwell-Bloch equations following the procedure described
in Sec. \ref{sec3}, and more explicit in the appendix of Ref. \cite{lopez1},
with the physical parameters of a $CO_2$ laser.
Calculations show that any choice of those physical parameters leading
to a negative value of $A_{2R}'$ leads also to a negative value of $A_{1R}'$.
Therefore, the conditions of a "strong" HC in a $CO_2$ laser cannot be
fulfilled. \par
The second possibility will be called "weak" heteroclinic connection.
This corresponds to a positive real part for all nonlinear coefficients
of Eqs. (\ref{eqfive1}-\ref{eqfive2}).
In such a case, the set of conditions (\ref{condition1}-\ref{condition2})
written above are equivalent to:
\begin{equation}
M \equiv Min_0 \ Min_1 \ Min_2 > 1, \label{eqfin}
\end{equation}
where
\begin{eqnarray}
Min_0 \equiv Min (\frac{C^r}{H^r}, \frac{B^r}{G^r}), \\
Min_1 \equiv Min (\frac{G^r}{F^r}, \frac{I^r}{D^r}), \\
Min_2 \equiv Min (\frac{D^r}{A^r}, \frac{E^r}{C^r}).
\end{eqnarray}
Calculations with realistic parameters under usual conditions in a
$CO_2$ laser with a perfect $O(2)$ symmetry show numerical
evidence that $M$ has a maximum value equal to $1$. Nevertheless,
symmetry arguments do not exclude the possibility of finding
heteroclinic cycles in modified setups, provided that condition
(\ref{eqfin}) be satisfied. \par

\section{Conclusions and discussion}
The conditions for the existence of heteroclinic cycles in a
$CO_2$ laser with a cylindrical symmetry have been obtained. From
general symmetry arguments we proved that this kind of cycles are
not possible when there are only three or four modes -arising from
a supercritical bifurcation- interacting into the cavity. \newline
At least five modes must interact to produce that kind of cycles.
In this case the relationships that the coefficients must satisfy
for having heteroclinic connections are found. Calculations for
$CO_2$ laser systems  show that those requirements can not be
fulfilled under usual conditions.
\newline Some modification in the nonlinear coupling of the
Maxwell-Bloch equations is required, such as introducing a
nonlinear passive device into the cavity that modifies the
nonlinear coefficients.

\section{Acknowledgements}
We would like to thank G.B. Mindlin (Buenos Aires), 
J.R. Tredicce (Nice) and F.T. Arecchi (Florence) for
very useful and fruitful discussions.

\newpage
\begin{center} {\bf Figure Captions} \end{center}\par
{\bf 1.} Hipotetic sketch of a possible heteroclinic connection
\{$Q_1$,$Q_2$,$Q_3$\} for a
3D-symmetric system of ODEs. Each line $Q_i(t)$ connects the fixed points ($P_i$,
$P_{i+1}$) and lies on the correspondent invariant plane $\Pi_i$.\par
{\bf 2.} 2D-projections of the flow in the amplitude space for the interaction
in a $CO_2$ laser of modes $0$ and $\pm 1$.\par
{\bf 3.} A fifth mode, $\rho_0$, connecting the two planes of symmetry,
$(\rho_1,\rho_2)-(\rho_3,\rho_4)$, of equations (\ref{eqn1}-\ref{eqn2})
is needed in order to build a heteroclinic connection.\par
{\bf 4.} A generic heteroclinic connection between the
rotating waves of two interacting modes.\par
{\bf 5.} Heteroclinic connection established with five interacting modes in
a $CO_2$ laser: $\rho_0 \rightarrow plane\:(\rho_3, \rho_4)\;
\rightarrow\; plane\:(\rho_1, \rho_2)\; \rightarrow\; \rho_0$.

\end{document}